\documentclass[aps,prl,twocolumn,superscriptaddress,showpacs,nobalancelastpage]{revtex4}

\newcommand{\sigphin}{$\sigma_{\phi N}$}
\newcommand{\sigphininel}{$\sigma_{\phi N}^{inel}$}
\newcommand{\gevc}{~GeV$^2$/c$^2$}

\usepackage{graphicx}

\begin{document}


\title{
First measurement of coherent $\phi$-meson photoproduction on deuteron at low energies}

\newcommand*{\OHIOU}{Ohio University, Athens, Ohio  45701}
\affiliation{\OHIOU}
\newcommand*{\DUKE}{Duke University, Durham, North Carolina 27708}
\affiliation{\DUKE}
\newcommand*{\JLAB}{Thomas Jefferson National Accelerator Facility, Newport News, Virginia 23606}
\affiliation{\JLAB}
\newcommand*{\SCAROLINA}{University of South Carolina, Columbia, South Carolina 29208}
\affiliation{\SCAROLINA}
\newcommand*{\ANL}{Argonne National Laboratory, Argonne, IL 60439}
\affiliation{\ANL}
\newcommand*{\ASU}{Arizona State University, Tempe, Arizona 85287-1504}
\affiliation{\ASU}
\newcommand*{\UCLA}{University of California at Los Angeles, Los Angeles, California  90095-1547}
\affiliation{\UCLA}
\newcommand*{\CSU}{California State University, Dominguez Hills, Carson, CA 90747}
\affiliation{\CSU}
\newcommand*{\CMU}{Carnegie Mellon University, Pittsburgh, Pennsylvania 15213}
\affiliation{\CMU}
\newcommand*{\CUA}{Catholic University of America, Washington, D.C. 20064}
\affiliation{\CUA}
\newcommand*{\SACLAY}{CEA-Saclay, Service de Physique Nucl\'eaire, 91191 Gif-sur-Yvette, France}
\affiliation{\SACLAY}
\newcommand*{\CNU}{Christopher Newport University, Newport News, Virginia 23606}
\affiliation{\CNU}
\newcommand*{\UCONN}{University of Connecticut, Storrs, Connecticut 06269}
\affiliation{\UCONN}
\newcommand*{\ECOSSEE}{Edinburgh University, Edinburgh EH9 3JZ, United Kingdom}
\affiliation{\ECOSSEE}
\newcommand*{\FU}{Fairfield University, Fairfield CT 06824}
\affiliation{\FU}
\newcommand*{\FIU}{Florida International University, Miami, Florida 33199}
\affiliation{\FIU}
\newcommand*{\FSU}{Florida State University, Tallahassee, Florida 32306}
\affiliation{\FSU}
\newcommand*{\GWU}{The George Washington University, Washington, DC 20052}
\affiliation{\GWU}
\newcommand*{\ECOSSEG}{University of Glasgow, Glasgow G12 8QQ, United Kingdom}
\affiliation{\ECOSSEG}
\newcommand*{\ISU}{Idaho State University, Pocatello, Idaho 83209}
\affiliation{\ISU}
\newcommand*{\INFNFR}{INFN, Laboratori Nazionali di Frascati, 00044 Frascati, Italy}
\affiliation{\INFNFR}
\newcommand*{\INFNGE}{INFN, Sezione di Genova, 16146 Genova, Italy}
\affiliation{\INFNGE}
\newcommand*{\ORSAY}{Institut de Physique Nucleaire ORSAY, Orsay, France}
\affiliation{\ORSAY}
\newcommand*{\ITEP}{Institute of Theoretical and Experimental Physics, Moscow, 117259, Russia}
\affiliation{\ITEP}
\newcommand*{\JMU}{James Madison University, Harrisonburg, Virginia 22807}
\affiliation{\JMU}
\newcommand*{\KYUNGPOOK}{Kyungpook National University, Daegu 702-701, Republic of Korea}
\affiliation{\KYUNGPOOK}
\newcommand*{\UMASS}{University of Massachusetts, Amherst, Massachusetts  01003}
\affiliation{\UMASS}
\newcommand*{\MOSCOW}{Moscow State University, General Nuclear Physics Institute, 119899 Moscow, Russia}
\affiliation{\MOSCOW}
\newcommand*{\UNH}{University of New Hampshire, Durham, New Hampshire 03824-3568}
\affiliation{\UNH}
\newcommand*{\NSU}{Norfolk State University, Norfolk, Virginia 23504}
\affiliation{\NSU}
\newcommand*{\ODU}{Old Dominion University, Norfolk, Virginia 23529}
\affiliation{\ODU}
\newcommand*{\RPI}{Rensselaer Polytechnic Institute, Troy, New York 12180-3590}
\affiliation{\RPI}
\newcommand*{\RICE}{Rice University, Houston, Texas 77005-1892}
\affiliation{\RICE}
\newcommand*{\URICH}{University of Richmond, Richmond, Virginia 23173}
\affiliation{\URICH}
\newcommand*{\UNIONC}{Union College, Schenectady, NY 12308}
\affiliation{\UNIONC}
\newcommand*{\VT}{Virginia Polytechnic Institute and State University, Blacksburg, Virginia   24061-0435}
\affiliation{\VT}
\newcommand*{\VIRGINIA}{University of Virginia, Charlottesville, Virginia 22901}
\affiliation{\VIRGINIA}
\newcommand*{\WM}{College of William and Mary, Williamsburg, Virginia 23187-8795}
\affiliation{\WM}
\newcommand*{\YEREVAN}{Yerevan Physics Institute, 375036 Yerevan, Armenia}
\affiliation{\YEREVAN}
\newcommand*{\NOWUNH}{University of New Hampshire, Durham, New Hampshire 03824-3568}
\newcommand*{\NOWUMASS}{University of Massachusetts, Amherst, Massachusetts  01003}
\newcommand*{\NOWMIT}{Massachusetts Institute of Technology, Cambridge, Massachusetts  02139-4307}
\newcommand*{\NOWCUA}{Catholic University of America, Washington, D.C. 20064}
\newcommand*{\NOWECOSSEE}{Edinburgh University, Edinburgh EH9 3JZ, United Kingdom}
\newcommand*{\NOWGEISSEN}{Physikalisches Institut der Universitaet Giessen, 35392 Giessen, Germany}
\newcommand*{\TRIUMF}{TRIUMF, 4004, Wesbrook Mall, Vancouver, BC, V6T 2A3, Canada}

\author {T.~Mibe}
\affiliation{\OHIOU}
\author {H.~Gao}
\affiliation{\DUKE}
\author {K.~Hicks}
\affiliation{\OHIOU}
\author {K.~Kramer}
\affiliation{\DUKE}
\author {S.~Stepanyan}
\affiliation{\JLAB}
\author {D.J.~Tedeschi}
\affiliation{\SCAROLINA}
\author {M.J.~Amaryan} 
\affiliation{\ODU}
\author {P.~Ambrozewicz} 
\affiliation{\FIU}
\author {M.~Anghinolfi} 
\affiliation{\INFNGE}
\author {G.~Asryan} 
\affiliation{\YEREVAN}
% ...
\collaboration{The CLAS Collaboration}
     \noaffiliation

\date{\today}

\begin{abstract}
The cross section and decay angular distributions
for the coherent $\phi$ meson photoproduction on the deuteron have been 
measured for the first time up to a squared four-momentum transfer 
$t = (p_{\gamma}-p_{\phi})^2 =-2$\gevc, 
using the CLAS detector at the Thomas Jefferson National Accelerator Facility. 
The cross sections are compared with predictions from a re-scattering model.
In a framework of vector meson dominance, the data 
are consistent with the total $\phi$-N cross section \sigphin ~at about 10~mb.
If vector meson dominance is violated, a larger \sigphin 
~is possible by introducing 
larger $t$-slope for the $\phi N\rightarrow\phi N$ process 
than that for the $\gamma N\rightarrow\phi N$ process.
The decay angular distributions of the $\phi$ 
are consistent with helicity conservation.
\end{abstract}


\maketitle


The exchange of gluons between hadrons, 
known as Pomeron exchange~\cite{Donnachie:2002},
is a fundamental 
process that is expected to dominate hadron-hadron total cross sections at 
high energies.
In general, multi-gluon exchange is harder to study at lower energy 
since diagrams including quark exchange play a more important role.
The $\phi$ meson is unique in that it is nearly pure $s\bar{s}$ 
and hence multi-gluon exchange is expected to dominate $\phi$-N 
scattering at all energies.  Since gluon exchange is flavor 
blind, information on multi-gluon exchange, isolated 
from the $\phi$-N interaction, would be universal and useful
in models of hadron-hadron interactions.
For example, information on the $\phi$-N interaction at very low energies,
known as the QCD van der Waals interaction,
is essential for the reliable prediction of the possible 
formation of a bound state in the $\phi$-N system~\cite{Gao:2000az}.

The total $\phi$-N cross section (\sigphin)
is estimated by using vector meson dominance (VMD) applied
to exclusive $\phi$ photoproduction on the proton 
in the photon energy range $E_{\gamma}< 10$~GeV,
resulting in \sigphin $\simeq$ 10--12~mb~\cite{Behrend:1978ik,Sibirtsev:2006yk},
which is in agreement with the estimate 
from the additive quark model~\cite{PhysRevLett.16.1015}
applied to $KN$ and $\pi N$ scattering data~\cite{vmreview}.
More recently, the inelastic $\phi$-N cross section \sigphininel ~was extracted from 
the attenuation of $\phi$-mesons in photoproduction from 
Li, C, Al, and Cu nuclei~\cite{Ishikawa:2004id}. 
The attenuation for large $A$ is significantly 
larger than that calculated from VMD.
More sophisticated models~\cite{Cabrera:2003wb,Muhlich:2005kf,Sibirtsev:2006yk} 
are consistent with the experiment if \sigphininel ~is significantly larger ($\sim$30~mb)
compared with \sigphin ~from the VMD model.
The reason for the discrepancy of \sigphin ~from these two estimates 
is not well understood. Here we will show that information on the $t$ dependence and
spin structure of the $\phi$-N interaction provides essential clues to solve this
problem.

\begin{figure}[tb]
\includegraphics[width=4.2cm]{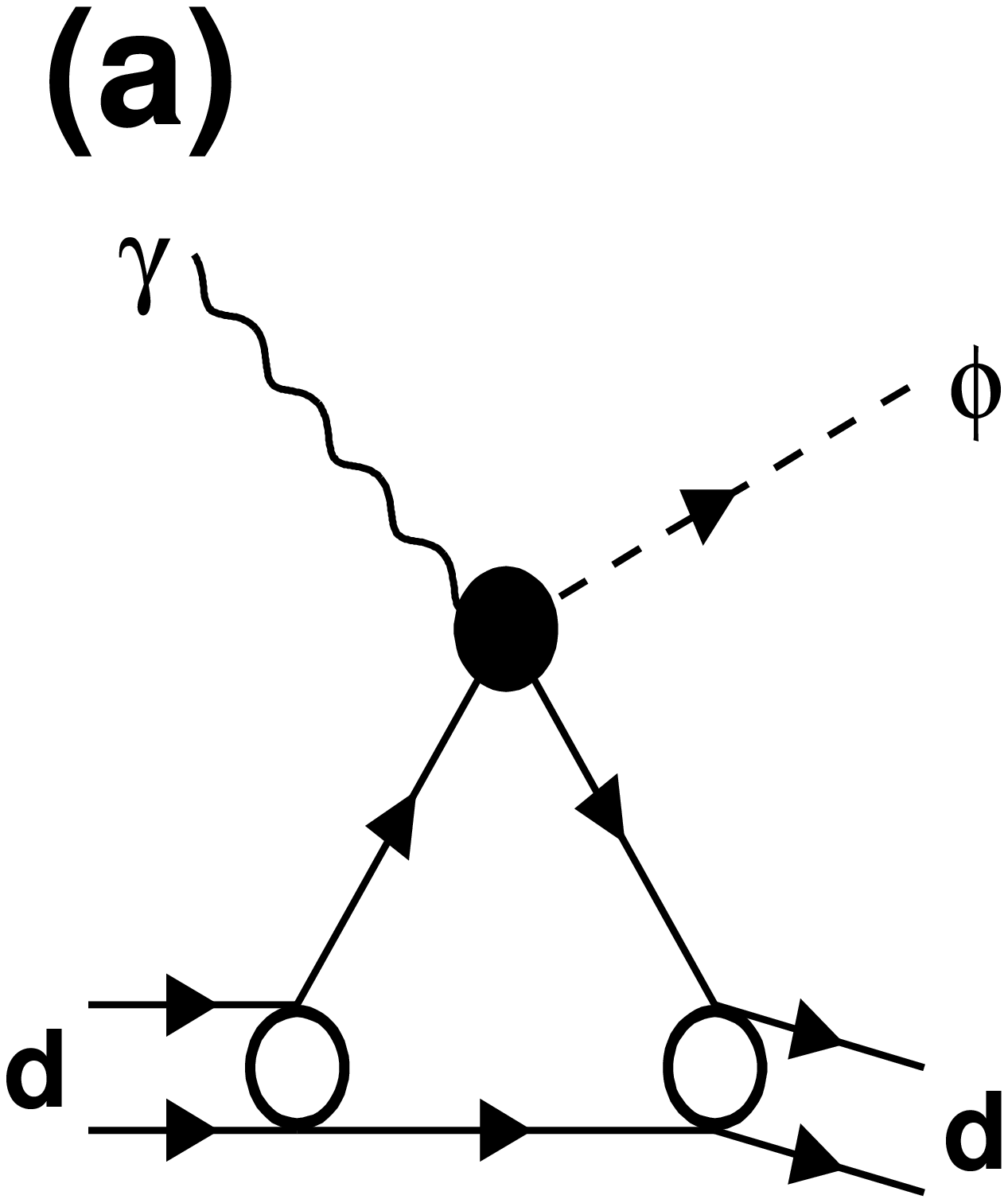}
\includegraphics[width=4.2cm]{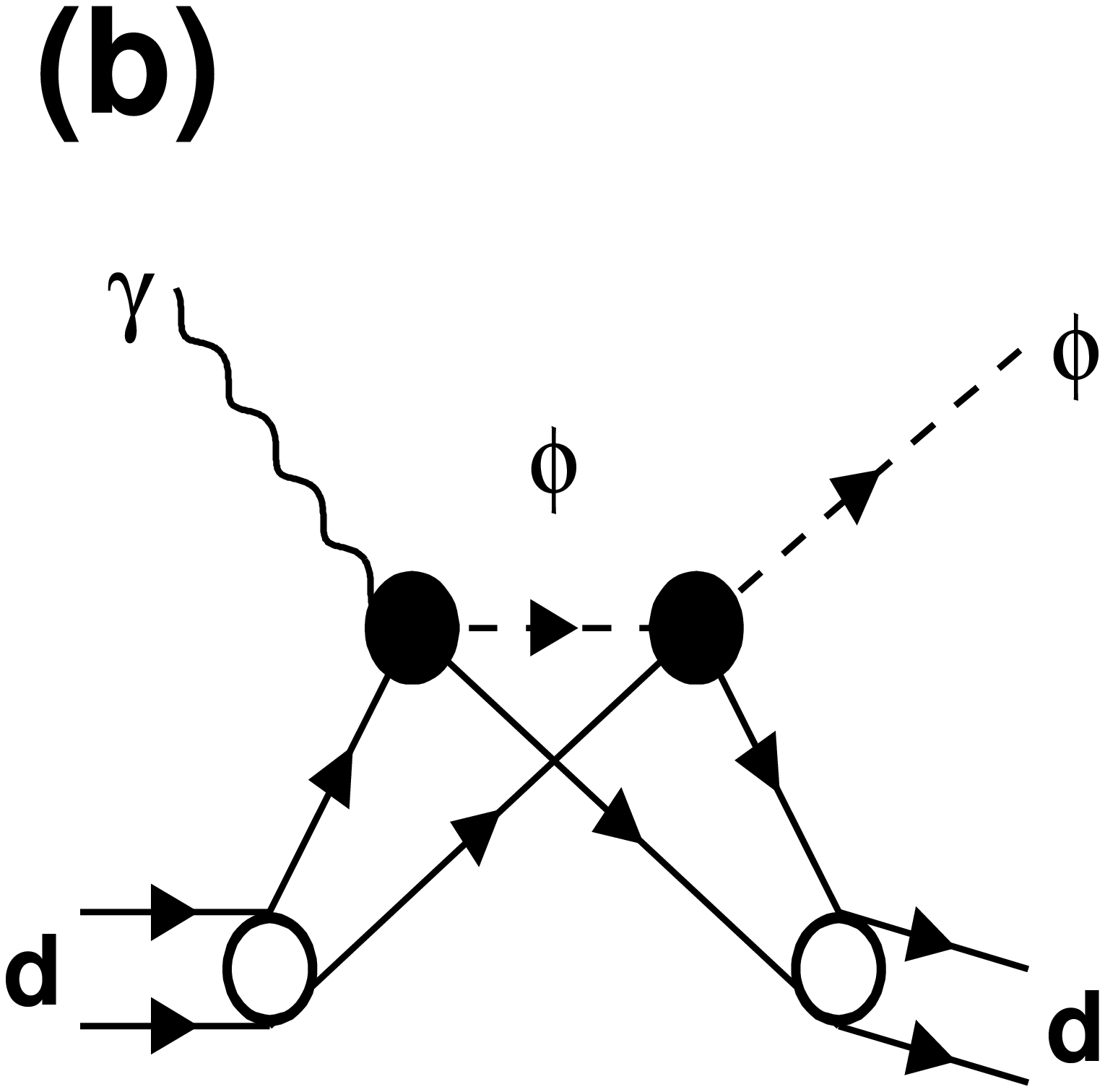}
\caption{(a) Single-scattering and (b) double-scattering contributions to
the coherent $\phi$-meson photoproduction on the deuteron.}
\label{fig:single-double}
\end{figure}

In this Letter, the $\phi$-N interaction is 
investigated in coherent photoproduction on deuterium.
The diagrams of the dominant processes contributing to the reaction
$\gamma d \rightarrow \phi d$ are shown 
in Fig.~\ref{fig:single-double}.  In the first diagram, Fig.~\ref{fig:single-double}(a), 
the $\phi$ is produced in a single scattering of a nucleon, 
which is dominant at small $-t$, and
strongly suppressed at larger $-t$
due to the deuteron form factor.
The second diagram, Fig.~\ref{fig:single-double}(b),
shows double-scattering, where the $\phi$ is produced at 
the first vertex and scatters from the other nucleon at 
the second vertex. The strength of the second interaction 
is gauged by \sigphin. The probability to undergo 
double-scattering increases
at larger $-t$ because both nucleons receive 
momentum transfer and may recombine into a final-state 
deuteron with a smaller relative momentum between the two nucleons~\cite{Frankfurt:1997ss}.

The $\phi$ meson is a spin one particle which decays to a $K\bar{K}$ pair, {\it i.e.} two spin-less
particles.
The decay angular distribution of the $\phi$ carries information on
the spin structure of the reaction amplitude which is the sum of single- and double-
scattering processes~\cite{Schilling:1970um}.

The measurement of the 
differential cross sections of coherent $\phi$ photoproduction 
and the decay angular distributions in a wide $t$ range
allows one to study the $\phi$-N interaction in both single and double scattering,
as well as the transition from one to the other.

The data were collected with the CLAS detector
and the Hall B tagged-photon beam
at the Thomas Jefferson National Accelerator Facility~\cite{Mecking:2003zu}.
The incident electron beam energy was 3.8~GeV, producing tagged photons
in the range from 0.8 to 3.6~GeV.
The photon beam was directed onto a 24-cm long liquid-deuterium target.
The data acquisition trigger required two charged particles detected 
in coincidence with a tagged photon.
Charged particles were momentum analyzed by the CLAS torus magnet and 
three sets of drift chambers. The torus magnet was run at
two settings, low field (2250~A) and high field (3375~A), each for 
about half of the run period.

\begin{figure}[t] 
\includegraphics[width=8.5cm]{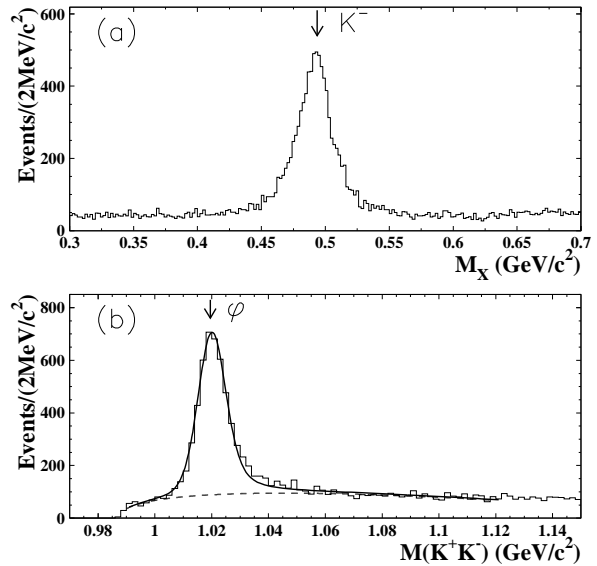}
\caption{ 
(a) Missing mass distribution of the reaction $\gamma d \rightarrow d K^+ X$
for events near the $\phi$-meson mass ($0.98<M(K^+K^-)<1.12$~GeV/c$^2$).
(b) Invariant mass distribution for the $K^+K^-$ pair 
after the selection of the missing $K^-$.
The solid curve is a fit to the data. The dashed curve shows the contribution
from background.}
\label{fig.mm} 
\end{figure}

The reaction $\gamma d \rightarrow \phi d$ was identified by detecting 
a deuteron and a $K^+$ from $\phi \rightarrow K^+K^-$ decay.
The $K^+$ and deuteron were selected based on time-of-flight,
path length, and momentum measurements. Figure~\ref{fig.mm}(a) shows the missing mass 
distribution, $M_X$, for the reaction $\gamma d \rightarrow dK^+X$
when events near the $\phi$-meson peak ($0.98<M(K^+K^-)<1.12$~GeV/c$^2$)
were selected in the $K^+K^-$ invariant mass, assuming a $K^-$ was the missing particle.
A missing $K^-$ peak is seen on top of a smooth background from non-$d$~$K^+K^-$ final 
states. 
The missing mass resolution, ranging from 8 to 30~MeV/c$^2$, 
depends on photon energy and the deuteron momentum.
A three-$\sigma$ cut was applied to select the missing $K^-$ 
for the exclusive $\gamma d \rightarrow K^+K^- d$ reaction.

Figure~\ref{fig.mm}(b) shows the invariant mass distribution for the $K^+K^-$ pair 
after the selection of the missing $K^-$.
The $\phi$-meson peak appears above a smooth background. 
The $\phi$-meson yield was obtained from a 
fit to the $M(K^+K^-)$ distribution by a gaussian-convoluted Breit-Wigner function
and a background function. The width and the pole position for the Breit-Wigner function
were fixed to 4.3~MeV/c$^2$ and 1019.5~MeV/c$^2$, respectively~\cite{Yao:2006px}.
The standard deviation of the gaussian distribution was fixed to the value obtained from simulation.
The background function was chosen as
$a \sqrt{x^2-(2m_K)^2} + b (x^2-(2m_K)^2)$~\cite{Lukashin:2001sh},
where $x$ is $M(K^+K^-)$, $m_K$ is the charged kaon mass, and
$a$ and $b$ are the fit parameters.
Three background functions:
a linear background, background from
non-resonant $K^+K^-d$ production, and $f_0$ photoproduction, were
studied as alternative choices. 
The background models for the non-resonant $K^+K^-d$ and $f_0$ photoproduction 
were parameterized by the differential cross section and photon-energy distribution 
of events in the sidebands of the $\phi$-meson peak.
The dependence of the yield on the background function,
fit range, and parameterization of the Breit-Wigner function
were studied. 
The extracted yield changes between 3\% and 9\% depending on the yield extraction procedures. 

\begin{figure}[t]
\includegraphics[width=9.0cm]{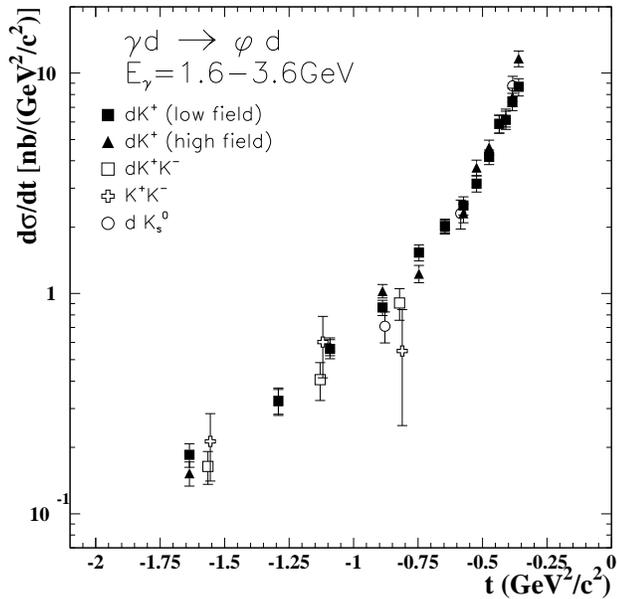}
\caption{Comparison of differential cross sections 
for $\gamma d \rightarrow \phi d$ from various topologies in the range $1.6<E_{\gamma}<3.6$~GeV.
Only statistical uncertainties are shown.
}
\label{fig.yieldalltopo2}
\end{figure}

The CLAS acceptance was determined by using 
a GEANT-based Monte Carlo simulation~\cite{Brun:1978fy}.
The simulation was iterated to reproduce measured $t$, 
photon energy, and decay angular distributions.
The acceptance was between 10\% and 20\% in the kinematic region 
covered by the present measurements.
The accuracy of the calculation of the acceptance was estimated from
comparison of results from the other event reconstruction topologies 
($d$~$K^+K^-$, $K^+K^-$, and $d$~$K^0_s$ topologies) for which
the acceptances were different from that for the $d$~$K^+$ topology.
The differential cross sections for these topologies are shown in
Fig.~\ref{fig.yieldalltopo2}. They agree with each other within statistical uncertainties,
indicating that the acceptance is understood as to 
the number of reconstructed tracks, charge combinations, and decay modes.
Supplemental simulations were performed to understand the systematic uncertainties due 
to the event generator (1-11\%) and event reconstruction (1-5\%).

Systematic uncertainties in the yield extraction and acceptance were estimated 
as a function of photon energy and $t$; they were between 4\% and 13\%.
The combined systematic uncertainty for 
the luminosity and trigger efficiency was less than 10\%.
Systematic uncertainties from contributions from 
accidental tracks, target windows, and particle misidentification 
are less than a few percent.
The total systematic uncertainty was estimated as 11-17\% by adding 
these uncertainties in quadrature.


\begin{table}[t]
\caption{Differential cross sections for the reaction $\gamma d \rightarrow \phi d$.
The second and third numbers in each field are the statistical and systematic uncertainties,
respectively.}
\label{tab.dsdt}
\begin{center}
\begin{tabular}{|c|c|r|r|}
\hline
\multicolumn{2}{|c|}{$t$ range (GeV$^2$/c$^2$)}& \multicolumn{2}{|c|}{$d\sigma/dt$~[nb/(GeV$^2$/c$^2$)]}\\
\hline
$t_{min}$ &$t_{max}$ & $1.6<E_{\gamma}<2.6$~GeV & $2.6<E_{\gamma}<3.6$~GeV\\
\hline
\hline
    -0.375 &   -0.350 &10.21 $\pm$ 0.82 (1.70) & 8.63 $\pm$ 0.80 (1.04)  \\
    -0.400 &   -0.375 & 8.85 $\pm$ 0.75 (1.11) & 6.80 $\pm$ 0.69 (1.07)  \\
    -0.425 &   -0.400 & 7.32 $\pm$ 0.59 (0.94) & 4.57 $\pm$ 0.53 (0.74)  \\
    -0.450 &   -0.425 & 6.16 $\pm$ 0.55 (0.81) & 5.76 $\pm$ 0.56 (0.65)  \\
    -0.500 &   -0.450 & 4.73 $\pm$ 0.34 (0.60) & 3.99 $\pm$ 0.33 (0.55)  \\
    -0.550 &   -0.500 & 3.52 $\pm$ 0.28 (0.51) & 3.59 $\pm$ 0.29 (0.55)  \\
    -0.600 &   -0.550 & 2.66 $\pm$ 0.24 (0.38) & 2.11 $\pm$ 0.22 (0.28)  \\
    -0.700 &   -0.600 & 2.17 $\pm$ 0.15 (0.26) & 1.83 $\pm$ 0.14 (0.24)  \\
    -0.800 &   -0.700 & 1.40 $\pm$ 0.12 (0.16) & 1.32 $\pm$ 0.12 (0.20)  \\
    -1.000 &   -0.800 & 0.94 $\pm$ 0.07 (0.11) & 0.96 $\pm$ 0.07 (0.11)  \\
    -1.200 &   -1.000 & 0.57 $\pm$ 0.06 (0.07) & 0.57 $\pm$ 0.05 (0.06)  \\
    -1.400 &   -1.200 & 0.28 $\pm$ 0.05 (0.04) & 0.36 $\pm$ 0.04 (0.05)  \\
    -2.000 &   -1.400 & 0.19 $\pm$ 0.02 (0.03) & 0.15 $\pm$ 0.02 (0.02)  \\
\hline

\end{tabular}
\end{center}
\end{table}

\begin{figure}[t] 
\includegraphics[width=9.0cm]{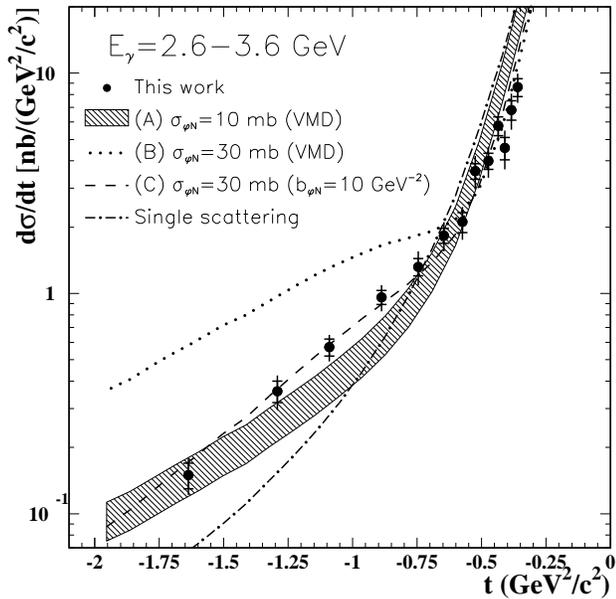}
\caption{Differential cross sections for the reaction $\gamma d \rightarrow \phi d$.
The inner error bars shown are statistical uncertainty only, while
the outer error bars are the sum of statistical and systematic uncertainties in quadrature.
The curves~A, B and C are calculations from 
the re-scattering model~\cite{Frankfurt:1997ss,Rogers:2005bt},
see text for details. 
The uncertainties on curves~B and C are comparable to that of curve~A,
but are not shown.
The dot-dashed curve is a contribution from 
the single scattering diagram.
}
\label{fig.cseg1a} 
\end{figure}

The differential cross sections were measured 
in the ranges $1.6<E_{\gamma}<2.6$~GeV and $2.6<E_{\gamma}<3.6$~GeV.
They are given in Table~\ref{tab.dsdt}. 
Figure~\ref{fig.cseg1a} shows the experimental data
in the range $2.6<E_{\gamma}<3.6$~GeV.
The data is compared with theoretical calculations 
using a rescattering model~\cite{Frankfurt:1997ss,Rogers:2005bt}.
In this model, 
the $\gamma N\rightarrow \phi N$ amplitude was parameterized by using 
published data on the $\gamma p\rightarrow \phi p$ reaction~\cite{Anciant:2000az}
and data from the proton target run during this experiment.
This amplitude was convoluted with the deuteron wave function 
with a correction of the relativistic-recoil effect~\cite{Frankfurt:1997ss}.
The double scattering process (Fig.~\ref{fig:single-double}(b)) is modeled
by the Generalized Eikonal Approximation \cite{Frankfurt:1996xx}.
The \sigphin ~and $t$ dependence for the re-scattering process
are the inputs for the calculation.
The model successfully 
reproduces the differential cross sections 
on coherent $\rho$ photoproduction~\cite{Frankfurt:1997ss} using the inputs from
the VMD.

The total model uncertainty is estimated to be about 20\%.
A 10\% uncertainty was assigned to the parametrization of 
the $\gamma N\rightarrow \phi N$ amplitude
based on the $\gamma p\rightarrow \phi p$ data. 
The effect of spin-flip 
in the process $\gamma N\rightarrow \phi N$ was ignored 
in the parametrization of the single scattering amplitude since 
the spin-flip amplitude is more suppressed in the coherent process than in the
incoherent process. 
A 15\% systematic uncertainty was assigned due to this effect~\cite{Titov:spinflip}.
An isospin dependence of the process $\gamma N\rightarrow \phi N$ was not taken into account
in the model, but Ref.~\cite{Titov:1998tx} suggests such an effect is small.

In Fig.~\ref{fig.cseg1a}, curve~A shows the $t$ distribution calculated by
using the VMD prediction for the $\phi$-N cross section, {\it i.e.} \sigphin=10~mb,
and the same $t$ distribution for the reaction $\gamma N\rightarrow \phi N$ 
and the reaction $\phi N\rightarrow \phi N$.
The curve~B corresponds to \sigphin=30~mb, inspired by Ref.~\cite{Ishikawa:2004id},
with the VMD assumption for the $t$ distribution.
It overestimates the data at large $-t$ where 
the contribution from double scattering dominates.
This implies that if the $t$ distribution follows the VMD prediction,
\sigphin ~should also be consistent with the VMD prediction.
In this case, inconsistency with the larger \sigphin 
~from the $A$-dependence experiment~\cite{Ishikawa:2004id} still remains.

However, the VMD picture may not be a good approximation in this photon energy range. 
The larger \sigphin ~from the $A$-dependence experiment~\cite{Ishikawa:2004id}
can be explained if the $t$ distribution of the reaction 
$\phi N \rightarrow \phi N$ is different from the VMD prediction.
For example, it is possible for the virtual $\phi$ to fluctuate to a $K \bar{K}$ pair 
and have a larger cross section for the second interaction~\cite{mark}.
In this case, the $t$-slope for the second interaction would be larger than
that for the $\gamma N\rightarrow \phi N$ reaction
based on a general geometric relation between the $t$-slope 
and the total cross section~\cite{Povh:1987ju}.
Following this hypothesis, cross sections were calculated
with \sigphin=30~mb using a larger exponential $t$-slope,
$b_{\phi N}=10$~(GeV/c)$^{-2}$, in the second interaction (curve~C).
The data is equally-well described by curve~C,
suggesting a larger $t$-slope parameter 
is necessary if \sigphin ~is larger than the VMD prediction.
Although the current data do not allow one to 
extract the \sigphin ~and the $t$-slope independently
due to the strong correlation between them, it possibly suggests 
a larger \sigphin ~from the $A$-dependence.

\begin{figure*}[t] 
\includegraphics[width=15.0cm]{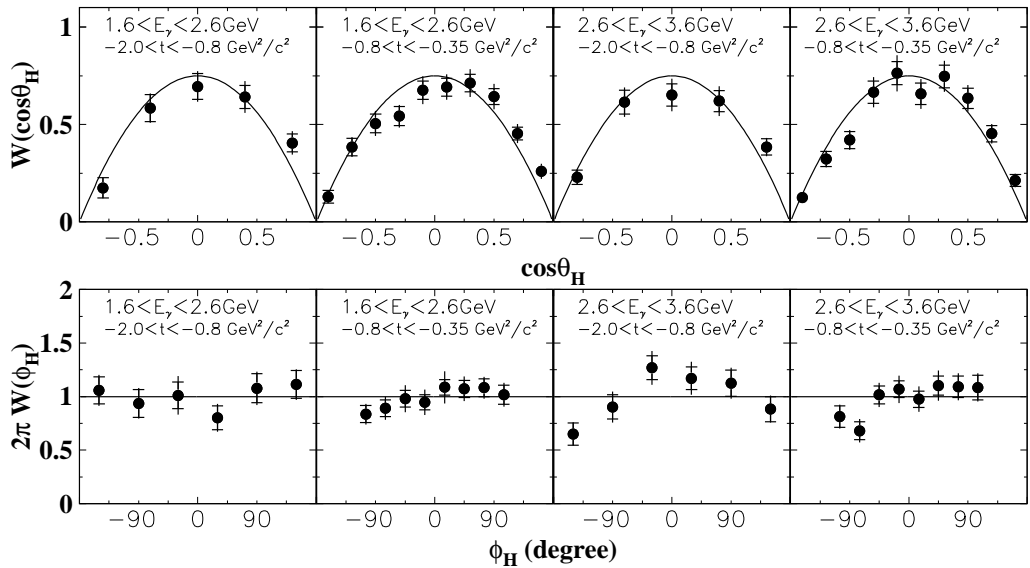}
\caption{Decay angular distributions of the $\phi$-meson 
in the helicity frame. The inner error bars shown are statistical uncertainty only, while
the outer error bars are the sum of statistical and systematic uncertainties in quadrature.
Solid curves are the predictions from helicity conservation.
}
\label{fig.w} 
\end{figure*}

In addition to the differential cross sections,
the decay angular distributions of the $\phi$ meson 
were also measured in the helicity frame~\cite{Schilling:1970um}.
The direction of the $\phi$-meson momentum in the CM system was chosen as the $z$-axis,
and the polar angle and azimuthal angle between the $K^+$ momentum and the $\phi$-meson production plane
were defined as $\theta_H$ and $\phi_H$ in the $\phi$-meson rest frame.
Figure~\ref{fig.w} shows the projections of the decay angular distributions 
onto $\cos\theta_H$ and $\phi_H$
in the ranges $-0.8<t<-0.35$~\gevc ~and $-2.0<t<-0.8$~\gevc ~in each photon energy region.
The data are consistent with the prediction from helicity conservation (solid curves),
{\it i.e.} the spin of the $\phi$-meson is aligned to the momentum of the $\phi$-meson.
This is similar to what was observed in the $\phi$ photoproduction 
on the proton~\cite{McCormick:2003bb,Mibe:2005er}.
In the larger $-t$ region, the double scattering contribution becomes more important. 
No drastic change is observed from the smaller $-t$ to the larger $-t$ region,
implying that the spin structure of the single- and double-scattering processes are similar.


In summary, we have presented the first measurement of 
the differential cross sections and 
decay angular distributions for 
coherent $\phi$ photoproduction on the deuteron up to $t=-2.0$\gevc.
The differential cross sections at large $-t$ exhibit a
contribution from double scattering.
The data are consistent with \sigphin =10~mb in a framework of VMD.
The data also provide a possible explanation for larger \sigphin ~if
the $t$-slope for $\phi N \rightarrow \phi N$ is larger than 
the VMD value from $\gamma p \rightarrow \phi p$.
The decay angular distributions follow the prediction from
helicity conservation.

This measurement demonstrates a new approach to the study of 
the $\phi$-N interaction in the low energy region
where VMD is not necessarily a good approximation.
Further measurements at higher photon energies~\cite{eg3},
at very small $-t$~\cite{lepscphi}, as well as
an $A$-dependence study in $e^+e^-$ decay~\cite{g7phi}
will make it possible
to map out details of the energy and $t$ dependences of the
$\phi$-N interaction.


We would like to thank the staff of the Accelerator and Physics
Divisions at Jefferson Lab who made this experiment possible.
We acknowledge useful discussions with T.~Rogers, M.~Sargsian,
M.~Strikman and A.~Titov. This work was supported in part by 
the Italian Istituto Nazionale de Fisica Nucleare, the French
Centre National de la Recherche Scientifique and Commissariat \`a
l'Energie Atomique, the Korea Research Foundation,
the U.S. Department of Energy and the
National Science Foundation, and the U.K. Engineering and
Physical Science Research Council.
Jefferson Science Associates (JSA) operates the
Thomas Jefferson National Accelerator Facility for the United States
Department of Energy under contract DE-AC05-06OR23177.

\bibliographystyle{apsrev}

\begin{thebibliography}{27}
\expandafter\ifx\csname natexlab\endcsname\relax\def\natexlab#1{#1}\fi
\expandafter\ifx\csname bibnamefont\endcsname\relax
  \def\bibnamefont#1{#1}\fi
\expandafter\ifx\csname bibfnamefont\endcsname\relax
  \def\bibfnamefont#1{#1}\fi
\expandafter\ifx\csname citenamefont\endcsname\relax
  \def\citenamefont#1{#1}\fi
\expandafter\ifx\csname url\endcsname\relax
  \def\url#1{\texttt{#1}}\fi
\expandafter\ifx\csname urlprefix\endcsname\relax\def\urlprefix{URL }\fi
\providecommand{\bibinfo}[2]{#2}
\providecommand{\eprint}[2][]{\url{#2}}

\bibitem[{\citenamefont{Donnachie et~al.}(2002)\citenamefont{Donnachie, Dosch,
  Landshoff, and Nachtmann}}]{Donnachie:2002}
\bibinfo{author}{\bibfnamefont{A.}~\bibnamefont{Donnachie}},
  \bibinfo{author}{\bibfnamefont{H.~G.} \bibnamefont{Dosch}},
  \bibinfo{author}{\bibfnamefont{P.~V.} \bibnamefont{Landshoff}},
  \bibnamefont{and}
  \bibinfo{author}{\bibfnamefont{O.}~\bibnamefont{Nachtmann}},
  \emph{\bibinfo{title}{Pomeron Physics and {QCD}}}
  (\bibinfo{publisher}{Cambridge University Press}, \bibinfo{year}{2002}).

\bibitem[{\citenamefont{Gao et~al.}(2001)\citenamefont{Gao, Lee, and
  Marinov}}]{Gao:2000az}
\bibinfo{author}{\bibfnamefont{H.}~\bibnamefont{Gao}},
  \bibinfo{author}{\bibfnamefont{T.~S.~H.} \bibnamefont{Lee}},
  \bibnamefont{and} \bibinfo{author}{\bibfnamefont{V.}~\bibnamefont{Marinov}},
  \bibinfo{journal}{Phys. Rev.} \textbf{\bibinfo{volume}{C63}},
  \bibinfo{pages}{022201(R)} (\bibinfo{year}{2001}).

\bibitem[{\citenamefont{Behrend et~al.}(1978)}]{Behrend:1978ik}
\bibinfo{author}{\bibfnamefont{H.~J.} \bibnamefont{Behrend}}
  \bibnamefont{et~al.}, \bibinfo{journal}{Nucl. Phys.}
  \textbf{\bibinfo{volume}{B144}}, \bibinfo{pages}{22} (\bibinfo{year}{1978}).

\bibitem[{\citenamefont{Sibirtsev et~al.}(2006)\citenamefont{Sibirtsev, Hammer,
  Meissner, and Thomas}}]{Sibirtsev:2006yk}
\bibinfo{author}{\bibfnamefont{A.}~\bibnamefont{Sibirtsev}},
  \bibinfo{author}{\bibfnamefont{H.~W.} \bibnamefont{Hammer}},
  \bibinfo{author}{\bibfnamefont{U.~G.} \bibnamefont{Meissner}},
  \bibnamefont{and} \bibinfo{author}{\bibfnamefont{A.~W.}
  \bibnamefont{Thomas}}, \bibinfo{journal}{Eur. Phys. J.}
  \textbf{\bibinfo{volume}{A 29}}, \bibinfo{pages}{209} (\bibinfo{year}{2006}).

\bibitem[{\citenamefont{Lipkin}(1966)}]{PhysRevLett.16.1015}
\bibinfo{author}{\bibfnamefont{H.~J.} \bibnamefont{Lipkin}},
  \bibinfo{journal}{Phys. Rev. Lett.} \textbf{\bibinfo{volume}{16}},
  \bibinfo{pages}{1015} (\bibinfo{year}{1966}).

\bibitem[{\citenamefont{Rosenfeld and S\"oding}(1973)}]{vmreview}
\bibinfo{author}{\bibfnamefont{A.~H.} \bibnamefont{Rosenfeld}}
  \bibnamefont{and} \bibinfo{author}{\bibfnamefont{P.}~\bibnamefont{S\"oding}},
  \emph{\bibinfo{title}{Properties of the Fundamental Interactions}}
  (\bibinfo{publisher}{Editrice Compositori, Bologna}, \bibinfo{year}{1973}),
  vol.~\bibinfo{volume}{9}, p. \bibinfo{pages}{882}.

\bibitem[{\citenamefont{Ishikawa et~al.}(2005)}]{Ishikawa:2004id}
\bibinfo{author}{\bibfnamefont{T.}~\bibnamefont{Ishikawa}}
  \bibnamefont{et~al.}, \bibinfo{journal}{Phys. Lett.}
  \textbf{\bibinfo{volume}{B608}}, \bibinfo{pages}{215} (\bibinfo{year}{2005}).

\bibitem[{\citenamefont{Cabrera et~al.}(2004)\citenamefont{Cabrera, Roca, Oset,
  Toki, and Vicente~Vacas}}]{Cabrera:2003wb}
\bibinfo{author}{\bibfnamefont{D.}~\bibnamefont{Cabrera}},
  \bibinfo{author}{\bibfnamefont{L.}~\bibnamefont{Roca}},
  \bibinfo{author}{\bibfnamefont{E.}~\bibnamefont{Oset}},
  \bibinfo{author}{\bibfnamefont{H.}~\bibnamefont{Toki}}, \bibnamefont{and}
  \bibinfo{author}{\bibfnamefont{M.~J.} \bibnamefont{Vicente~Vacas}},
  \bibinfo{journal}{Nucl. Phys.} \textbf{\bibinfo{volume}{A733}},
  \bibinfo{pages}{130} (\bibinfo{year}{2004}).

\bibitem[{\citenamefont{Muhlich and Mosel}(2006)}]{Muhlich:2005kf}
\bibinfo{author}{\bibfnamefont{P.}~\bibnamefont{Muhlich}} \bibnamefont{and}
  \bibinfo{author}{\bibfnamefont{U.}~\bibnamefont{Mosel}},
  \bibinfo{journal}{Nucl. Phys.} \textbf{\bibinfo{volume}{A765}},
  \bibinfo{pages}{188} (\bibinfo{year}{2006}).

\bibitem[{\citenamefont{Frankfurt
  et~al.}(1997{\natexlab{a}})}]{Frankfurt:1997ss}
\bibinfo{author}{\bibfnamefont{L.}~\bibnamefont{Frankfurt}}
  \bibnamefont{et~al.}, \bibinfo{journal}{Nucl. Phys.}
  \textbf{\bibinfo{volume}{A622}}, \bibinfo{pages}{511}
  (\bibinfo{year}{1997}{\natexlab{a}}).

\bibitem[{\citenamefont{Schilling et~al.}(1970)\citenamefont{Schilling,
  Seyboth, and Wolf}}]{Schilling:1970um}
\bibinfo{author}{\bibfnamefont{K.}~\bibnamefont{Schilling}},
  \bibinfo{author}{\bibfnamefont{P.}~\bibnamefont{Seyboth}}, \bibnamefont{and}
  \bibinfo{author}{\bibfnamefont{G.~E.} \bibnamefont{Wolf}},
  \bibinfo{journal}{Nucl. Phys.} \textbf{\bibinfo{volume}{B15}},
  \bibinfo{pages}{397} (\bibinfo{year}{1970}).

\bibitem[{\citenamefont{Mecking et~al.}(2003)}]{Mecking:2003zu}
\bibinfo{author}{\bibfnamefont{B.~A.} \bibnamefont{Mecking}}
  \bibnamefont{et~al.}, \bibinfo{journal}{Nucl. Instrum. Meth.}
  \textbf{\bibinfo{volume}{A503}}, \bibinfo{pages}{513} (\bibinfo{year}{2003}).

\bibitem[{\citenamefont{Yao et~al.}(2006)}]{Yao:2006px}
\bibinfo{author}{\bibfnamefont{W.~M.} \bibnamefont{Yao}} \bibnamefont{et~al.}
  (\bibinfo{collaboration}{{P}article {D}ata {G}roup}), \bibinfo{journal}{J.
  Phys.} \textbf{\bibinfo{volume}{G33}}, \bibinfo{pages}{1}
  (\bibinfo{year}{2006}).

\bibitem[{\citenamefont{Lukashin et~al.}(2001)}]{Lukashin:2001sh}
\bibinfo{author}{\bibfnamefont{K.}~\bibnamefont{Lukashin}}
  \bibnamefont{et~al.}, \bibinfo{journal}{Phys. Rev.}
  \textbf{\bibinfo{volume}{C63}}, \bibinfo{pages}{065205}
  (\bibinfo{year}{2001}).

\bibitem[{\citenamefont{Brun et~al.}(1978)\citenamefont{Brun, Hagelberg,
  Hansroul, and Lassalle}}]{Brun:1978fy}
\bibinfo{author}{\bibfnamefont{R.}~\bibnamefont{Brun}},
  \bibinfo{author}{\bibfnamefont{R.}~\bibnamefont{Hagelberg}},
  \bibinfo{author}{\bibfnamefont{M.}~\bibnamefont{Hansroul}}, \bibnamefont{and}
  \bibinfo{author}{\bibfnamefont{J.~C.} \bibnamefont{Lassalle}}
  (\bibinfo{year}{1978}), \bibinfo{note}{{C}ERN-DD-78-2-REV}.

\bibitem[{\citenamefont{Rogers et~al.}(2006)\citenamefont{Rogers, Sargsian, and
  Strikman}}]{Rogers:2005bt}
\bibinfo{author}{\bibfnamefont{T.~C.} \bibnamefont{Rogers}},
  \bibinfo{author}{\bibfnamefont{M.~M.} \bibnamefont{Sargsian}},
  \bibnamefont{and} \bibinfo{author}{\bibfnamefont{M.~I.}
  \bibnamefont{Strikman}}, \bibinfo{journal}{Phys. Rev.}
  \textbf{\bibinfo{volume}{C73}}, \bibinfo{pages}{045202}
  (\bibinfo{year}{2006}).

\bibitem[{\citenamefont{Anciant et~al.}(2000)}]{Anciant:2000az}
\bibinfo{author}{\bibfnamefont{E.}~\bibnamefont{Anciant}} \bibnamefont{et~al.},
  \bibinfo{journal}{Phys. Rev. Lett.} \textbf{\bibinfo{volume}{85}},
  \bibinfo{pages}{4682} (\bibinfo{year}{2000}).

\bibitem[{\citenamefont{Frankfurt
  et~al.}(1997{\natexlab{b}})\citenamefont{Frankfurt, Sargsian, and
  Strikman}}]{Frankfurt:1996xx}
\bibinfo{author}{\bibfnamefont{L.~L.} \bibnamefont{Frankfurt}},
  \bibinfo{author}{\bibfnamefont{M.~M.} \bibnamefont{Sargsian}},
  \bibnamefont{and} \bibinfo{author}{\bibfnamefont{M.~I.}
  \bibnamefont{Strikman}}, \bibinfo{journal}{Phys. Rev.}
  \textbf{\bibinfo{volume}{C56}}, \bibinfo{pages}{1124}
  (\bibinfo{year}{1997}{\natexlab{b}}).

\bibitem[{\citenamefont{Titov}()}]{Titov:spinflip}
\bibinfo{author}{\bibfnamefont{A.~I.} \bibnamefont{Titov}},
  \bibinfo{note}{private communications}.

\bibitem[{\citenamefont{Titov et~al.}(1999)\citenamefont{Titov, Lee, and
  Toki}}]{Titov:1998tx}
\bibinfo{author}{\bibfnamefont{A.~I.} \bibnamefont{Titov}},
  \bibinfo{author}{\bibfnamefont{T.~S.~H.} \bibnamefont{Lee}},
  \bibnamefont{and} \bibinfo{author}{\bibfnamefont{H.}~\bibnamefont{Toki}},
  \bibinfo{journal}{Phys. Rev.} \textbf{\bibinfo{volume}{C59}},
  \bibinfo{pages}{R2993} (\bibinfo{year}{1999}).

\bibitem[{\citenamefont{Rogers et~al.}()\citenamefont{Rogers, Sargsian, and
  Strikman}}]{mark}
\bibinfo{author}{\bibfnamefont{T.~C.} \bibnamefont{Rogers}},
  \bibinfo{author}{\bibfnamefont{M.~M.} \bibnamefont{Sargsian}},
  \bibnamefont{and} \bibinfo{author}{\bibfnamefont{M.}~\bibnamefont{Strikman}},
  \bibinfo{note}{private communications}.

\bibitem[{\citenamefont{Povh and Hufner}(1987)}]{Povh:1987ju}
\bibinfo{author}{\bibfnamefont{B.}~\bibnamefont{Povh}} \bibnamefont{and}
  \bibinfo{author}{\bibfnamefont{J.}~\bibnamefont{Hufner}},
  \bibinfo{journal}{Phys. Rev. Lett.} \textbf{\bibinfo{volume}{58}},
  \bibinfo{pages}{1612} (\bibinfo{year}{1987}).

\bibitem[{\citenamefont{McCormick et~al.}(2004)}]{McCormick:2003bb}
\bibinfo{author}{\bibfnamefont{K.}~\bibnamefont{McCormick}}
  \bibnamefont{et~al.}, \bibinfo{journal}{Phys. Rev.}
  \textbf{\bibinfo{volume}{C69}}, \bibinfo{pages}{032203}
  (\bibinfo{year}{2004}).

\bibitem[{\citenamefont{Mibe et~al.}(2005)}]{Mibe:2005er}
\bibinfo{author}{\bibfnamefont{T.}~\bibnamefont{Mibe}} \bibnamefont{et~al.},
  \bibinfo{journal}{Phys. Rev. Lett.} \textbf{\bibinfo{volume}{95}},
  \bibinfo{pages}{182001} (\bibinfo{year}{2005}).

\bibitem[{\citenamefont{Gothe et~al.}()\citenamefont{Gothe, Holtrop, Smith,
  Stepanyan et~al.}}]{eg3}
\bibinfo{author}{\bibfnamefont{R.}~\bibnamefont{Gothe}},
  \bibinfo{author}{\bibfnamefont{M.}~\bibnamefont{Holtrop}},
  \bibinfo{author}{\bibfnamefont{E.}~\bibnamefont{Smith}},
  \bibinfo{author}{\bibfnamefont{S.}~\bibnamefont{Stepanyan}},
  \bibnamefont{et~al.}, \bibinfo{note}{{J}efferson Laboratory experiment
  {E}04-010, PAC25 proposal}.

\bibitem[{\citenamefont{Chang et~al.}()}]{lepscphi}
\bibinfo{author}{\bibfnamefont{W.~C.} \bibnamefont{Chang}}
  \bibnamefont{et~al.}, \bibinfo{note}{private communications}.

\bibitem[{\citenamefont{Wood et~al.}(2006)}]{g7phi}
\bibinfo{author}{\bibfnamefont{M.~H.} \bibnamefont{Wood}} \bibnamefont{et~al.}
  (\bibinfo{year}{2006}), \bibinfo{note}{{C}LAS {A}pproved {A}nalysis
  proposal}.

\end{thebibliography}

\end{document}